\documentclass[superscriptaddress,showpacs,amssymb,10pt,reprint,aps,prd,longbibliography,nofootinbib,floatfix]{revtex4-2}

\usepackage{graphicx,epsfig,amssymb,times} 
\usepackage{amsmath,amsfonts}
\usepackage{bm}
\usepackage{epstopdf}
\usepackage{hyperref}
\usepackage[caption=false]{subfig}
\usepackage[usenames]{color}   
\usepackage[dvipsnames]{xcolor}
\usepackage[normalem]{ulem}
\usepackage{multirow}
 
\definecolor{coolblack}{rgb}{0.0, 0.18, 0.39}
\definecolor{darkred}{rgb}{0.5,0,0}
\definecolor{darkgreen}{rgb}{0,0.5,0}
\definecolor{darkblue}{rgb}{0,0,0.5}
\definecolor{lapislazuli}{rgb}{0.15, 0.38, 0.61}
\definecolor{venetianred}{rgb}{0.78, 0.03, 0.08}
\definecolor{bleudefrance}{rgb}{0.19, 0.55, 0.91}
\definecolor{dogwoodrose}{rgb}{0.84, 0.09, 0.41}
\hypersetup{colorlinks=true, citecolor=darkblue, linkcolor=darkblue, 
urlcolor = darkblue}

\begin{document}
\title{\large Geodesic analysis, absorption and scattering in the static Hayward spacetime}
	
\author{Marco A. A. de Paula}
\email{marco.paula@icen.ufpa.br}
\affiliation{Programa de P\'os-Gradua\c{c}\~{a}o em F\'{\i}sica, Universidade 
		Federal do Par\'a, 66075-110, Bel\'em, Par\'a, Brazil.}

\author{Luiz C. S. Leite}
\email{luiz.leite@ifpa.edu.br}
\affiliation{Campus Altamira, Instituto Federal do Par\'a, 68377-630, Altamira, Par\'a, Brazil.}		
	
\author{Lu\'is C. B. Crispino}
\email{crispino@ufpa.br}
\affiliation{Programa de P\'os-Gradua\c{c}\~{a}o em F\'{\i}sica, Universidade 
		Federal do Par\'a, 66075-110, Bel\'em, Par\'a, Brazil.}
	\affiliation{Departamento de Matem\'atica da Universidade de Aveiro and Centre for Research and Development in Mathematics and Applications (CIDMA), Campus de Santiago, 3810-183 Aveiro, Portugal.}

\begin{abstract}

We investigate the propagation of massless particles and scalar fields in the background of Hayward regular black holes. We compute the absorption and scattering cross sections and compare our numerical results with some analytical approximations, showing that they are in excellent agreement. We show that some of the absorption and scattering results of Reissner-Nordstr\"om black holes can be mimicked by Hayward regular black holes, for appropriate choices of the charge of the two different black holes.

\end{abstract}

\date{\today}

\maketitle

\section{Introduction}

In recent years, experiments testing the strong-field regime have consolidated general relativity (GR) as a robust theory to describe gravity~\cite{AAA2016,AAA2019L1,AAA2022L12,GA2023}. Despite its achievements, GR also predicts the existence of curvature singularities in the core of the standard black hole (BH) solutions. We may argue that the limitations of Einstein's Theory at the BH center rely on its classical formalism. Therefore, a fully quantum gravity theory, which would successfully combine GR and quantum field theory, could avoid the formation of curvature singularities. Yet, although there have been efforts in this direction (see, e.g., Ref.~\cite{CPB2004} for a review), a fully successful quantum gravity theory has not been obtained so far.

As an alternative to the standard BH solutions of GR, there are the so-called regular BH (RBH) spacetimes, i.e. (curvature) singularity-free BH geometries. Nonsingular static spacetimes can be obtained by requiring an effective cutoff in the energy density at the BH center, preventing the spacetime metric from diverging at $r = 0$. This can be accomplished by demanding that the spacetime behaves as a de Sitter~\cite{B1968,D1992-2,B1994,ABG1998,D2004,H2006} or Minkowski~\cite{BV2014,C2015,SV2019} geometry at the BH core. (For reviews on RBHs and possible physical sources see Refs.~\cite{A2008,DPS2021,SZ2022,CL2023}.)

The Hayward geometry~\cite{H2006} is an example of spacetime which can have no curvature singularities, avoid the mass inflation phenomena~\cite{BKS2021} (at least from the classical point of view), and can be interpreted as a BH solution sourced by a nonlinear magnetic monopole~\cite{FW2016,F2017,TSA2018,MA2018}~\footnote{The Reissner-Nordstr\"om solution can also be regarded as a BH solution sourced by a magnetic monopole~\cite{C2019}.}. Due to these features, the Hayward geometry has gained a lot of attention over the past few years. We can improve our understanding of the Hayward spacetime by investigating how it interacts with surrounding fields. In this context, we can compute for Hayward BHs the absorption and scattering cross sections [a subject which have been extensively studied since the 1960s in several BH scenarios (see, e.g., Refs.~\cite{FHM1988,DDL2006,COM2007,CDE2009,OCH2011,CB2014,MC2014,CDHO2014,MOC2015,BC2016,S2017,SBP2018,AD2019,JBC2020,MLC2020-2,PLC2020,PLC2022,JBC2022,PLC2023c,MLC2020-3,SX2023} and references therein)].

We investigate the absorption and scattering properties of neutral massless test scalar fields in the background of Hayward RBHs. The remainder of this paper is organized as follows. In Sec.~\ref{sec:hs}, we introduce the Hayward spacetime as a solution of GR minimally coupled to nonlinear electrodynamics (NED). In Sec.~\ref{sec:ga}, we investigate the trajectories of massless particles and also consider the semiclassical glory approximation. The partial wave-analysis is applied in Sec.~\ref{sec:pwa} to obtain the absorption and scattering cross sections of the massless scalar field. In Sec.~\ref{sec:mr}, we present our main results concerning the absorption and scattering of neutral massless test scalar fields in Hayward spacetime. Our concluding remarks are stated in Sec.~\ref{sec:fr}. Throughout this work, we consider natural units, for which $G = c = \hbar = 1$, and signature $+2$.

\section{Hayward spacetime}\label{sec:hs}

The action associated with the minimal coupling between GR and NED can be written as
\begin{equation}
\label{action}\mathcal{S} = \dfrac{1}{16\pi}\int d^{4}x\left(R-\mathcal{L}(F)\right)\sqrt{-g},
\end{equation}
where $R$ is the Ricci scalar, $\mathcal{L}(F)$ is a gauge-invariant Lagrangian density, and $g$ is the determinant of the metric tensor $g_{\mu\nu}$. The function $F$ is the Maxwell scalar, namely
\begin{equation}
\label{MS}F = F_{\mu\nu}F^{\mu\nu},
\end{equation}
with $F_{\mu\nu}$ being the standard electromagnetic field tensor. By varying the action~\eqref{action} with respect to $g_{\mu\nu}$, we get
\begin{equation}
\label{fieldequations}G_{\mu}^{\ \nu} = T_{\mu}^{\ \nu} = 2\left(\mathcal{L}_{F}F_{\mu\sigma}F^{\nu\sigma} -\dfrac{1}{4}\delta_{\mu}^{\ \nu}\mathcal{L}(F)\right),
\end{equation}
in which $\mathcal{L}_{F} \equiv \partial \mathcal{L}/\partial F$. The dynamic field equations of the electromagnetic field are given by
\begin{equation}
\label{EFTC}\nabla_{\mu}\left(\mathcal{L}_{F}F^{\mu\nu}\right)  = \ 0 \ \ \ \text{and} \ \ \  \nabla_{\mu}\star F^{\mu\nu} = 0,
\end{equation}  
where $\star F^{\mu\nu}$ is the dual electromagnetic field tensor.

We consider a static and spherically symmetric line element given by
\begin{equation}
\label{D_LE}ds^{2} = -f(r)dt^{2}+f(r)^{-1}dr^{2}+r^{2}d\Omega^{2},
\end{equation}
where $f(r)$ is the metric function to be determined by the field equations~\eqref{fieldequations} and $d\Omega^{2} = d\theta^{2}+\sin^{2}\theta d\varphi^{2}$ is the line element of a $2$-dimensional unit sphere. In this context, the only non-null components of the electromagnetic field tensor are given by $F_{23} = -F_{32} = Q\sin\theta$, so that the Maxwell scalar is
\begin{equation}
\label{MS_HL}F = \dfrac{2Q^{2}}{r^{4}}.
\end{equation}

\begin{figure}[!htbp]
\begin{centering}
    \includegraphics[width=\columnwidth]{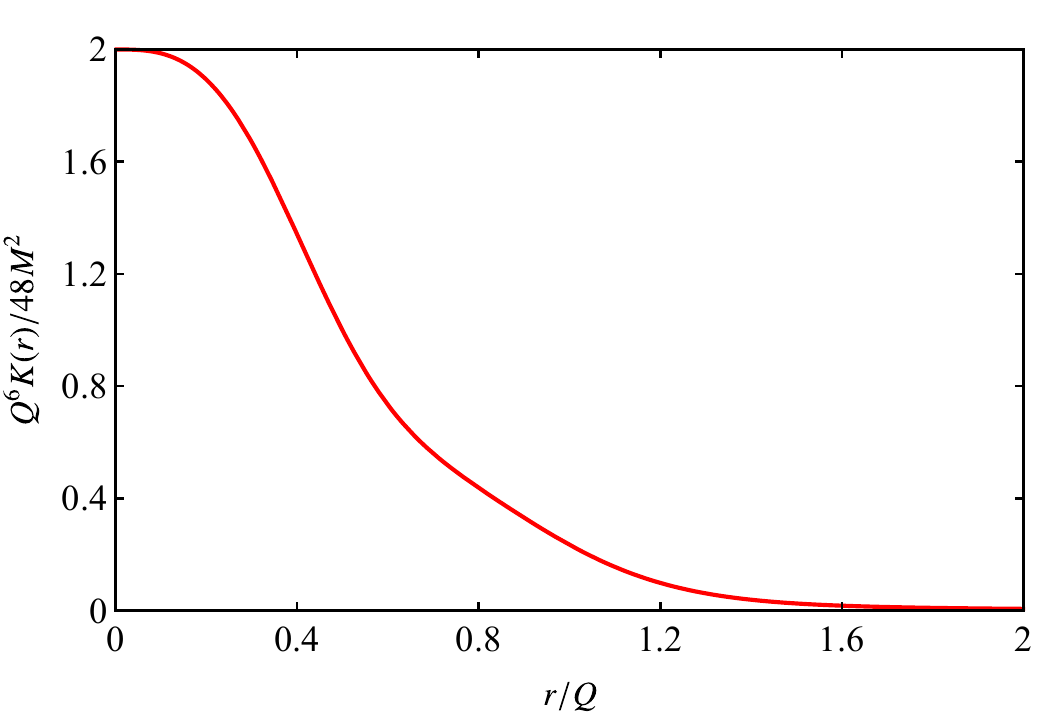}
    \caption{Kretschmann scalar invariant of the Hayward spacetime, as a function of $r/Q$. In the limit $r \rightarrow 0$ the scalar is finite and non null, namely $K(r)|_{r = 0} = 96M^{2}/Q^{6}$.}
    \label{kscalar}
\end{centering}
\end{figure}

The NED model associated with the Hayward spacetime can be written as~\cite{FW2016,F2017,MA2018}
\begin{equation}
\label{LD_HL}\mathcal{L}(F) = \dfrac{12M}{|Q|Q^{2}}\dfrac{\left(Q^{2} F/2 \right)^{\frac{3}{2}}}{\left(1+\left( Q^{2} F/2\right)^{\frac{3}{4}} \right)^{2}},
\end{equation}
where $Q$ and $M$ are the magnetic charge and mass of the central object, respectively. By using the equation $G_{0}^{\ 0} = T_{0}^{\ 0}$, we obtain the Hayward metric function
\begin{equation}
\label{MF_HD}f(r) = 1-\dfrac{2Mr^{2}}{r^{3}+Q^{3}}.
\end{equation}
In the chargeless limit $(Q \rightarrow 0)$, the Hayward metric function reduces to the Schwarzschild one. In Fig.~\ref{kscalar}, we display the Kretschmann scalar invariant, given by
\begin{equation}
K = R_{\mu\nu\sigma\rho}R^{\mu\nu\sigma\rho},
\end{equation}
where $R_{\mu\nu\sigma\rho}$ is the Riemann tensor, for the Hayward spacetime. Throughout this paper, we consider $Q > 0$, what is sufficient to guarantee the absence of curvature singularities for $r \geq 0$~\cite{BR2013,ZM2023}.

The line element~\eqref{D_LE}, considering the metric function~\eqref{MF_HD}, describes RBHs when the condition $Q \leq Q_{\rm{ext}} \approx 1.0582M$ is satisfied, where $Q_{\rm{ext}}$ is the extreme charge value. We can obtain $Q_{\rm{ext}}$ by solving $f(r) = 0$ and $f^{\prime}(r) = 0$ simultaneously, where $\prime$ denotes a differentiation with respect to the radial coordinate $r$. For $Q < Q_{\rm{ext}}$, we have two horizons, given by the real roots of $f(r) = 0$. We denote the Cauchy horizon and the event horizon as $r_{-}$ and $r_{+}$, respectively. For $Q = Q_{\rm{ext}}$, the two horizons coincide.
For its turn, $Q > Q_{\rm{ext}}$ leads to horizonless solutions, which is beyond the scope of this work. Moreover, we exhibit our results in terms of the normalized charge, defined as
\begin{equation}
\alpha \equiv \dfrac{Q}{Q_{\rm{ext}}},
\end{equation}
which facilitates comparisons between different spacetimes. In the Reissner-Nordstr\"om (RN) case, $Q$ can represent an electric or magnetic charge, while in the Hayward geometry $Q$ can be identified exclusively as a magnetic charge~\cite{FW2016,F2017}.

For large $r$, the Hayward metric behaves as
\begin{equation}
\label{Asy_a}f(r) = 1-\dfrac{2M}{r} + \dfrac{2MQ^{3}}{r^{4}} + \mathcal{O}\left[\dfrac{1}{r^{5}}\right],
\end{equation}
whereas at the core we have
\begin{equation}
f(r) = 1 - \dfrac{2M}{Q^{3}}r^{2} + \mathcal{O}[r]^{5}.
\end{equation}
Therefore, the spacetime is asymptotically flat as $r \rightarrow \infty$ and has a de Sitter behavior at the center. As occurs in the Bardeen geometry~\cite{B1968,ABG2000}, the Hayward spacetime does not satisfy a correspondence with the Maxwell theory for large $r$ since the corresponding NED model~\eqref{LD_HL} does not behave as $\mathcal{L}(F) \rightarrow F$ for small $F$. We also point out that the behavior of the Hayward geometry at its center is a common feature of NED-based RBHs that satisfy the weak energy condition~\cite{D2004}. In this context, the energy density of the NED source is maximal and finite at the solution core, preventing it from diverging as $r \rightarrow 0$, in contrast with linear electrodynamics. 

\section{Geodesic analysis}\label{sec:ga}

In this section, we investigate the propagation of massless particles in Hayward RBH spacetimes. Due to the spherical symmetry of the geometry, we treat the equations of motion in the equatorial plane, i.e., $\theta = \pi /2$, without loss of generality.  We recall that in NED models, photons follow null geodesics of an effective metric tensor~\cite{P1970,GDP1981,MN2000}. Therefore, the classical results discussed in this section apply only for massless particles with nature other than electromagnetic.

\subsection{Trajectory of massless particles}

\begin{figure}[!htbp]
\begin{centering}
    \includegraphics[width=\columnwidth]{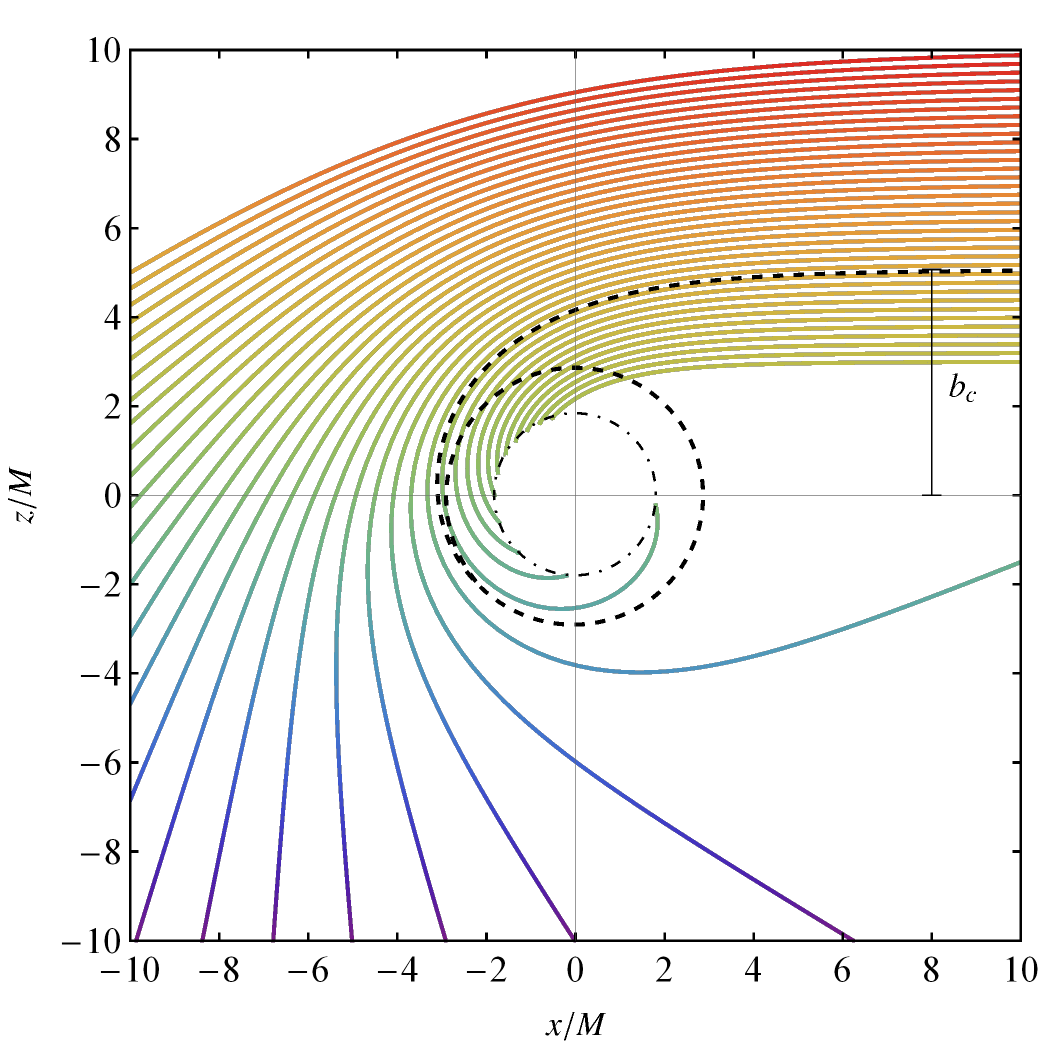}
    \caption{Trajectories of massless particles (with nature other than electromagnetic) in a Hayward spacetime with $\alpha = 0.8$, considering distinct impact parameters. The dashed curve is the trajectory associated with the corresponding critical impact parameter, given by $b_{c} = 5.0685M$, which ends up in an unstable circular orbit with $r_{c} = 2.8484$. The inner dot-dashed circle is the corresponding event horizon location, namely $r_{+} = 1.816M$. The initial conditions are given by $r_{\rm{inf}} = 100M$ and $\varphi = \pi - \arctan(3\sqrt{3}/100)$ at $t_{0} = 0$.}
    \label{geodesics}
\end{centering}
\end{figure}

The classical Lagrangian $\mathrm{L}$ that provides the equations of motion of particles in the spacetime~\eqref{D_LE} is given by
\begin{equation}
\label{L_SG}\mathrm{L} = \dfrac{1}{2} g_{\mu\nu}\dot{x}^{\mu}\dot{x}^{\nu},
\end{equation}
where the overdot corresponds to a differentiation with respect to an affine parameter. For massless particles, we have $\mathrm{L} = 0$. The constants of motion associated with $\mathrm{L}$ are given by
\begin{align}
\label{eqm1}E = f(r)\dot{t} \ \ \  \text{and} \ \ \ L = r^{2}\dot{\varphi},
\end{align}
where $E$ and $L$ are the energy and angular momentum of the particle, respectively. Using Eqs.~\eqref{L_SG}-\eqref{eqm1}, and the condition $\mathrm{L} = 0$, we may find a radial equation for massless particles given by
\begin{equation}
\label{ME}\dfrac{\dot{r}^{2}}{L^{2}} = V(r) = \dfrac{1}{b^{2}} - \dfrac{f(r)}{r^{2}},
\end{equation}
where $b \equiv L/E$ is the impact parameter. From $\dot{r}|_{r = r_{\rm{c}}} = 0$ and $\ddot{r}\big|_{r = r_{\rm{c}}} = 0$, we may find the critical radius $r_{c}$ of the unstable circular orbit and the critical impact parameter $b_{\rm{c}}$, namely
\begin{align}
\label{CR}2f(r_{\rm{c}})-&r_{\rm{c}}f'(r_{\rm{c}}) = 0,\\
\label{CIP}b_{\rm{c}} = \dfrac{L_{\rm{c}}}{E_{\rm{c}}} &= \dfrac{r_{\rm{c}}}{\sqrt{f(r_{\rm{c}})}},
\end{align}
respectively. In Fig.~\ref{geodesics}, we exhibit some geodesics of massless particles in the background of the Hayward spacetime. We can obtain these trajectories by numerically integrating the radial equation~\eqref{ME} and its first derivative. As we can observe, for $b < b_{c}$, the geodesics are absorbed, while for $b > b_{c}$ they are scattered. For its turn, the situation $b = b_{c}$ is related to a geodesic going round the BH in an unstable circular orbit (with radius $r_{c}$).

\begin{figure}[!htbp]
\begin{centering}
    \includegraphics[width=\columnwidth]{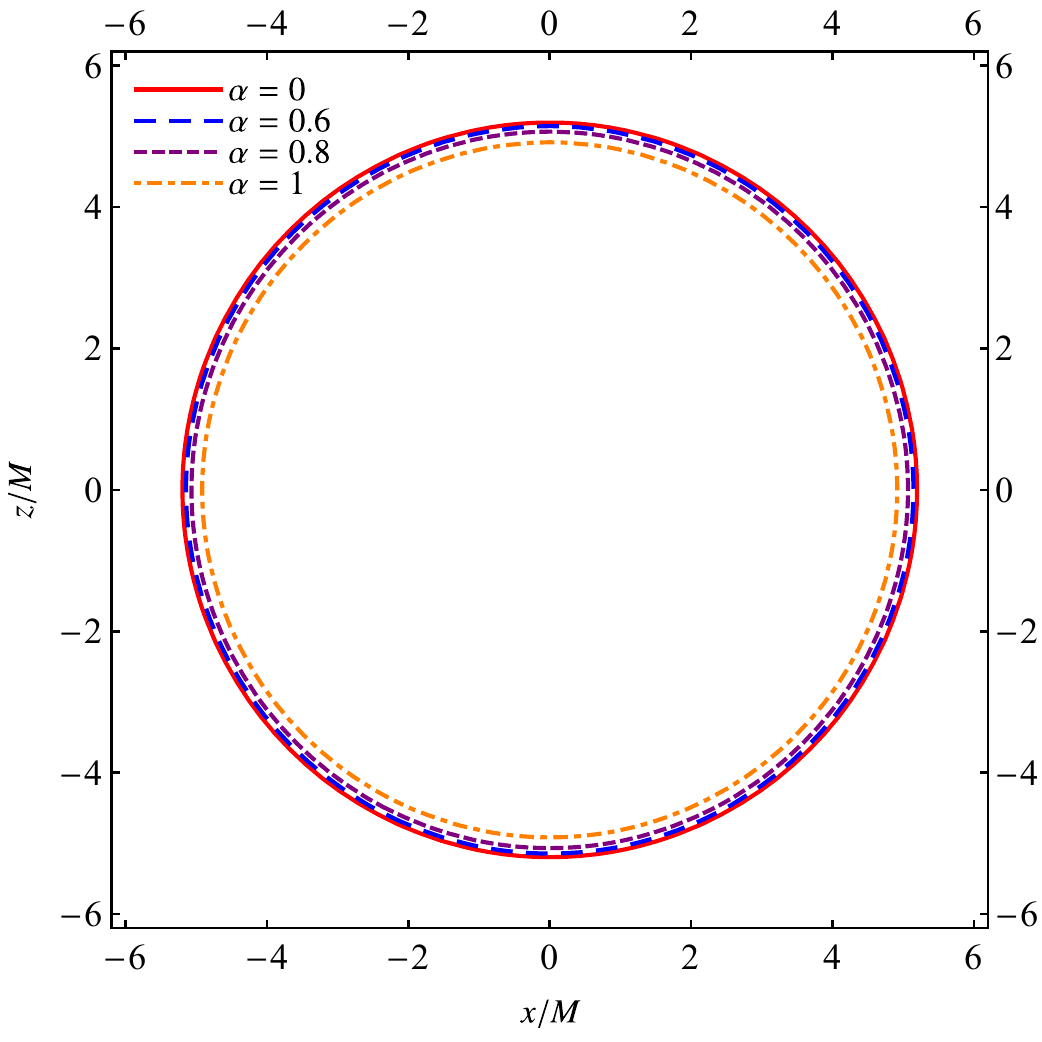}
    \caption{Shadows of the Hayward geometry, considering massless particles with nature other than electromagnetic, for distinct values of $\alpha$. We also consider the Schwarzschild case $\alpha = 0$, for comparison.}
    \label{shadowshd}
\end{centering}
\end{figure}

The classical capture cross section of geodesics, also known as geometric cross section (GCS), is given by~\cite{W1984}
\begin{equation}
\label{GCS}\sigma_{\rm{gcs}} \equiv \pi b_{\rm{c}}^{2}.
\end{equation}
Notice that in the spherically symmetric scenario, the critical impact parameter corresponds to the shadow radius as seen by a distant observer~\cite{CH2018,MP2023}. Therefore, the shape of the shadow can be obtained by the parametric plot of Eq.~\eqref{GCS}. This is illustrated in Fig.~\ref{shadowshd} where we exhibit the shadows of the Hayward spacetime, considering different values of $\alpha$ (and massless particles with nature other than electromagnetic).

At high energies, the absorption cross section (ACS) can be described by a formula known as the sinc approximation, which involves the GCS and the features of null unstable geodesics given by~\cite{S1978,DEF2011}
\begin{equation}
\label{SINC}\sigma_{\rm{hf}} \approx \sigma_{\rm{gcs}}\left[1-8\pi b_{c} \Lambda e^{-\pi b_{c} \Lambda} \text{sinc}(2\pi b_{c} \omega)\right],
\end{equation}
where $\text{sinc}(x) \equiv \sin(x)/x$ and $\Lambda$ is the Lyapunov exponent related to circular null geodesics~\cite{VC2009}, namely
\begin{equation}
\label{LEUCO}\Lambda = \sqrt{\dfrac{L^{2}_{c}}{2\dot{t}^{2}}\left(\dfrac{d^{2}V(r)}{dr^{2}}\right)\bigg|_{r = r_{c}}}.
\end{equation}

\subsection{Deflection angle in the weak-field limit}\label{subsec:dawfl}

By using the geodesic method, we can obtain an expression for the deflection angle and classical differential SCS in the weak-field limit. The turning point $r_0$, defined as the radius of maximum approximation of the (massless) particle, for a given value of $b$, satisfies the condition $\mathcal{U}(r)|_{r = r_{0}} = 0$, where
\begin{equation}
\label{U(r)}\mathcal{U}(r) \equiv \dfrac{dr}{d \varphi} = r^{2}\sqrt{\dfrac{1}{{b^2}}-\dfrac{f(r)}{r^{2}}}.
\end{equation}
Thus, the deflection angle of the scattered massless particle can be written as~\cite{N2013}
\begin{equation}
\label{DA}\Theta(b) = 2\int_{r_{0}}^{\infty} \dfrac{1}{\sqrt{\mathcal{U}(r)}}dr - \pi.
\end{equation} 

We can obtain an analytic expression for the deflection angle in the weak field limit by expanding the integrand of Eq.~\eqref{DA} in powers of $1/r$. The radius $r_{0}$ as a function of $b$ is obtained by solving Eq.~\eqref{CIP} and expanding the results in powers of $2M/b$. Following these steps, we can find that the weak deflection angle of massless particles in the background of Hayward and RN spacetimes, which are given by
\begin{align}
\label{thetaHD}\Theta(b)_{\rm{H}} & = \dfrac{4M}{b} + \dfrac{15\pi M^{2}}{4b^{2}}+\mathcal{O}\left[\dfrac{1}{b} \right]^{3}, \\
\label{thetaRN}\Theta(b)_{\rm{RN}} & = \dfrac{4M}{b} + \dfrac{3\pi\left(5M^{2}-Q^{2} \right)}{4b^{2}}+\mathcal{O}\left[\dfrac{1}{b} \right]^{3},
\end{align}
respectively~\footnote{From now on, we will abbreviate Hayward to H in equations and figures, whenever convenient.}. We see that the charge contributions do not modify the dominant term. Moreover, it can be shown that the charge contributions in the Hayward case will appear only for orders higher than $1/b^{3}$, due to the asymptotic behavior of the Hayward geometry [cf. Eq.~\eqref{Asy_a}]. 

The classical differential SCS is given by~\cite{N2013}
\begin{equation}
\label{CSCS}\dfrac{d\sigma_{\rm{cl}}}{d\Omega} = \dfrac{b}{\sin\theta}\bigg|\dfrac{db}{d\Theta}\bigg|,
\end{equation}
where $\theta$ is the scattering angle, which is related to the deflection angle by $\Theta=\theta-2n\pi$, with $n\in\mathbb{Z}^{+}$ being the number of times that the massless particle orbits the BH before being scattered to infinity. The classical SCS may be obtained by inverting Eq.~\eqref{DA} and inserting $b(\Theta)$ into Eq.~\eqref{CSCS}. In the weak field limit, we can use Eqs.~\eqref{thetaHD}-\eqref{thetaRN} to obtain the classical differential SCS for small scattering angles, which can be expressed as
\begin{align}
\label{CSCSweakHD}\dfrac{d\sigma_{\rm{cl}}^{\rm{H}}}{d\Omega} & = \dfrac{16M^2}{\Theta^4}+\dfrac{15\pi M^2}{4 \Theta ^3} + \mathcal{O}\left[\dfrac{1}{\Theta} \right]^{2}, \\
\label{CSCSweakRN}\dfrac{d\sigma_{\rm{cl}}^{\rm{RN}}}{d\Omega} & = \dfrac{16M^2}{\Theta ^4}+\dfrac{3\pi\left(5M^{2}-Q^{2}\right)}{4 \Theta ^3}+ \mathcal{O}\left[\dfrac{1}{\Theta} \right]^{2}.
\end{align}
Similarly to the weak deflection angle, the BH charge does not affect the leading term of the classical differential SCS.

\subsection{Semiclassical glory}

The semiclassical glory approximation can be used to unveil some wave scattering properties near the backward direction, i.e., $\theta = \pi$. In the background of a static and spherically symmetric BH geometry, the glory approximation for scalar waves can be written as~\cite{RAM1985}
\begin{equation}
\label{glory}\dfrac{d\sigma_{\rm{g}}}{d\Omega} = 2\pi \omega b_{g}^{2}\bigg|\dfrac{db}{d\theta}\bigg|_{\theta = \pi}J_{0}^{2}(\omega b_{g}\sin\theta),
\end{equation}
where $\omega$ is the frequency of the scalar wave, $b_{g}$ is the impact parameter of backscattered rays, and $J_{0}$ is the Bessel function of the first kind. Note that exist numerous values of $b_g$ corresponding to multiple values of $\theta=\Theta+2\pi n$, that result on backscattered null rays. The contributions to the glory scattering come from all the rays scattered near to $\theta\approx 180^{\circ}$. We know that the main contributions for the glory scattering are provided by the mode $n = 0$~\cite{CDE2009,MOC2015,PLC2022}. Therefore, we consider only $n = 0$ in the computation of the glory approximation.

\section{Partial-wave analysis}\label{sec:pwa}

In this section, we present the equation that governs the propagation of neutral massless test scalar fields in the background of the setup introduced in Sec.~\ref{sec:hs}. We also exhibit the differential SCS and total ACS of massless scalar waves in the background of spherically symmetric BHs.

\subsection{Massless scalar field}\label{subsec:msf}

The Klein-Gordon equation that governs the propagation of the neutral massless test scalar field $\Phi$ in the background of curved spacetimes reads
\begin{equation}
\label{KG}\dfrac{1}{\sqrt{-g}}\partial_{\mu}\left(\sqrt{-g}g^{\mu \nu}\partial_{\nu}\Phi\right) = 0.
\end{equation}
Within spherical symmetry, we can decompose $\Phi$ as
\begin{equation}
\label{PHI}\Phi= \dfrac{1}{r}\sum_{l}^{\infty} C_{\omega l}\Psi_{\omega l}(r)P_{l}(\cos\theta)e^{-i\omega t},
\end{equation}
where $C_{\omega l}$ are coefficients, with $\omega$ and $l$ being the frequency and angular momentum of the scalar field, respectively. The function $P_{l}$ is the Legendre polynomial and $\Psi_{\omega l}$ is the radial function. Using the tortoise coordinate $r_{\star}$, defined as $f(r)dr_{\star}=dr$, we can show that $\Psi_{\omega l}$ satisfies 
\begin{equation}
\label{RE_TC}\frac{d^{2}}{dr_{\star}^{2}}\Psi_{\omega l}+\left(\omega^{2}-V_{\rm{eff}}(r)\right)\Psi_{\omega l}=0,
\end{equation}
where the effective potential $V_{\rm{eff}}(r)$ is
\begin{equation}
\label{EP}V_{\rm{eff}}(r) = f(r)\left(\dfrac{1}{r}\dfrac{df(r)}{dr}+\dfrac{l(l+1)}{r^{2}}\right).
\end{equation}

\begin{figure}[!htbp]
\begin{centering}
    \includegraphics[width=1.0\columnwidth]{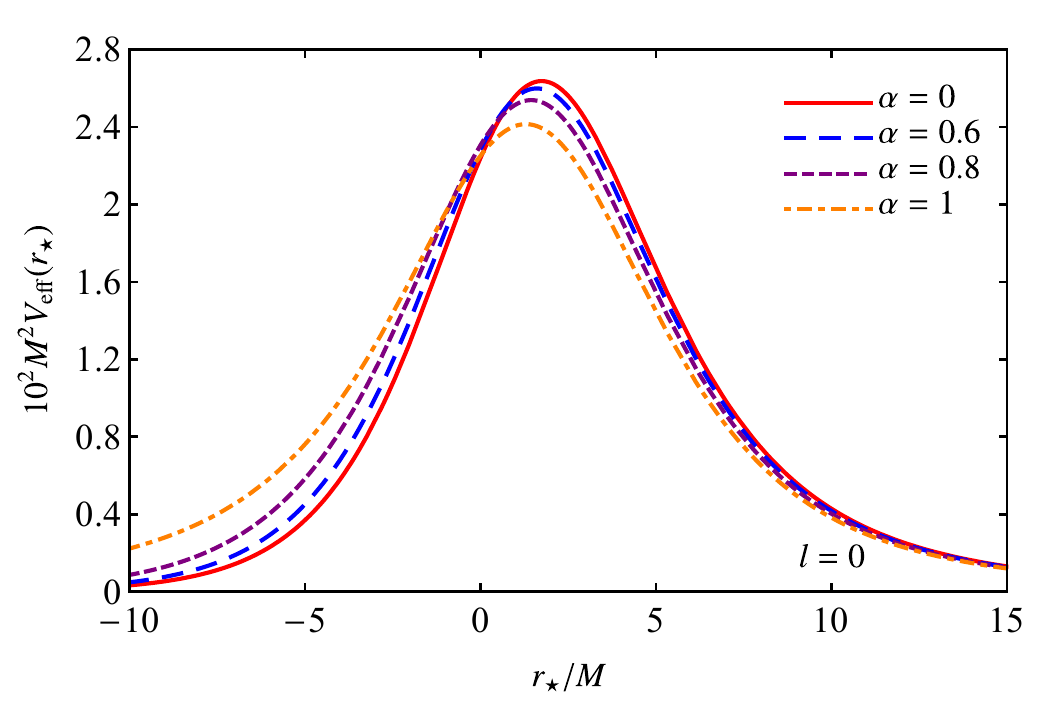}
    \includegraphics[width=1.0\columnwidth]{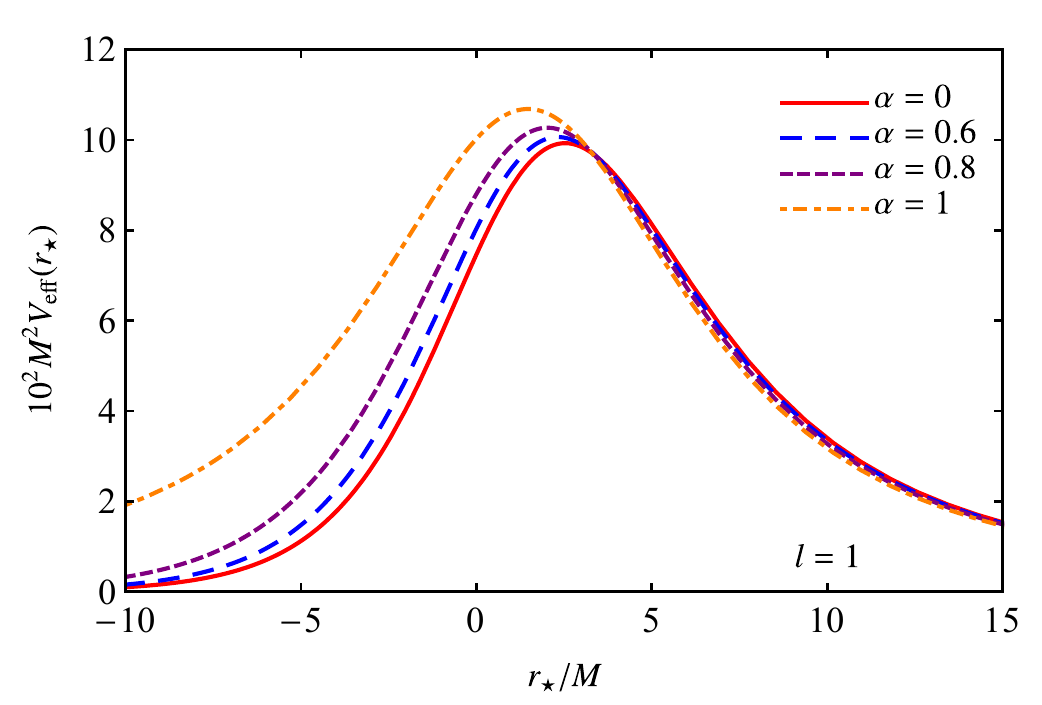}
    \caption{Effective potential of massless test scalar fields in the Hayward spacetime, as a function of $r_{\star}/M$, considering distinct values of $\alpha$, for $l = 0$ (top panel) and $l = 1$ (bottom panel).}
    \label{effecdifq}
\end{centering}
\end{figure}

In Fig.~\ref{effecdifq}, we display the effective potential in the Hayward spacetime for distinct values of $\alpha$ and $l$. Notice that for $l = 0$, the peak of the effective potential decreases as we increase the $\alpha$ values. However, for $l \geq 1$, the peak presents the opposite behavior. The behavior of the effective potential for $l = 0$ is remarkably different from the well-known (regular) BH solutions. 

\begin{figure}[!htbp]
\begin{centering}
    \includegraphics[width=1.0\columnwidth]{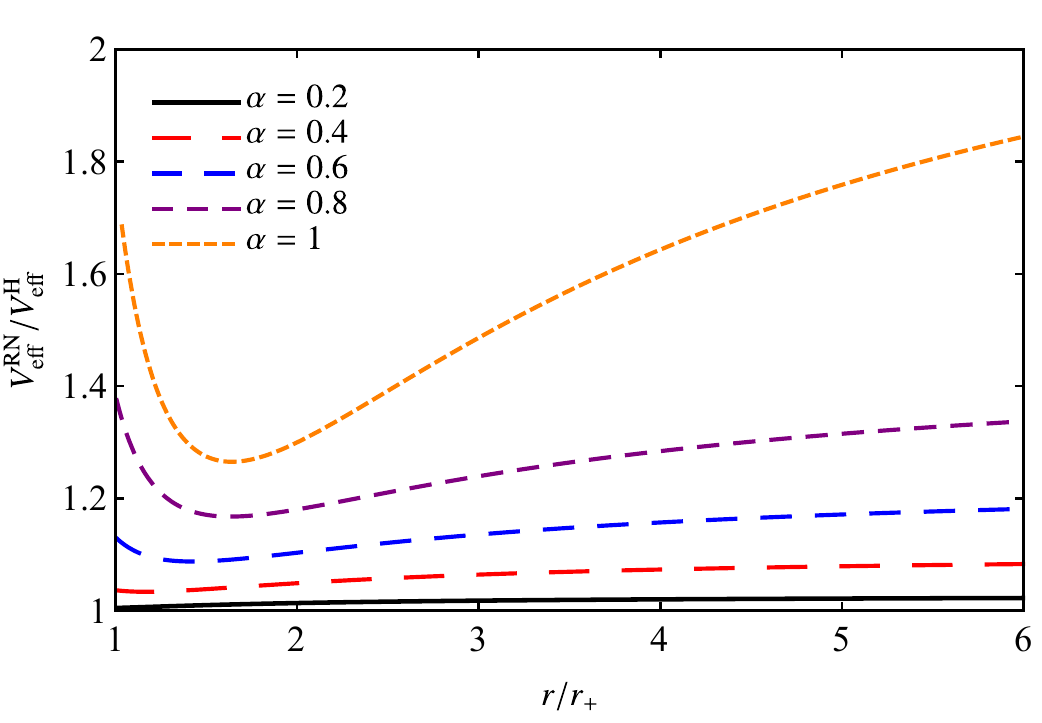}
    \caption{Ratio between the effective potential of massless test scalar fields in RN and Hayward spacetimes, as a function of $r/r_{+}$, for distinct values of $\alpha$. Here we consider $l = 0$.}
    \label{effecpot}
\end{centering}
\end{figure}

In Fig.~\ref{effecpot}, we compare the effective potentials in Hayward and RN spacetimes for fixed values of $\alpha$ and $l$, showing that, outside the event horizon, they typically satisfy
\begin{equation}
\label{Ratio}V_{\rm{eff}}^{\rm{RN}} > V_{\rm{eff}}^{\rm{H}}.
\end{equation}

The solutions of the Klein-Gordon equation consistent with the absorption/scattering problem are given by
\begin{equation}
\label{BC}\Psi_{\omega l}\sim\begin{cases}
T_{\omega l}e^{-i\omega r_{\star}}, & r_{\star}\rightarrow -\infty  \ (r\rightarrow r_{+}),\\
e^{-i\omega r_{\star}}+R_{\omega l}e^{i\omega r_{\star}}, & r_{\star}\rightarrow \infty \ (r\rightarrow \infty),
\end{cases}
\end{equation}
where $T_{\omega l}$ and $R_{\omega l}$ are complex coefficients, which satisfy
\begin{equation}
\label{CF}|R_{\omega l}|^{2}+|T_{\omega l}|^{2} = 1.
\end{equation}

\subsection{Absorption and scattering cross sections}

It is usual to obtain an expression for the total ACS $\sigma$ as a sum of partial waves contributions $\sigma_{l}$. For that purpose, we expand the scalar field as a sum of asymptotic plane waves and fix $C_{\omega l}$ with appropriated boundary conditions~\cite{U1976,CB2014}, resulting in
\begin{equation}
\label{ACS}\sigma = \sum_{l = 0}\sigma_{l},
\end{equation}
where $\sigma_{l}$ is given by
\begin{equation}
\label{PACS}\sigma_{l} = \dfrac{\pi}{\omega^{2}}(2l+1)\left(1-|R_{\omega l}|^{2}\right).
\end{equation}
For its turn, the differential SCS for static and spherically symmetric spacetimes can be written as~\cite{FHM1988}
\begin{equation}
\label{SCS}\dfrac{d\sigma}{d\Omega} = |h(\theta)|^{2},
\end{equation}
where $h(\theta)$ is the scattering amplitude given by
\begin{equation}
\label{scatta}h(\theta) = \dfrac{1}{2i\omega}\sum_{l = 0}^{\infty}(2l+1)[e^{2i\delta_{l}(\omega)}-1]P_{l}(\cos\theta),
\end{equation}
with the phase shifts $e^{2i\delta_{l}(\omega)}$ being defined as 
\begin{equation}
\label{ps}e^{2i\delta_{l}(\omega)} \equiv (-1)^{l+1}R_{\omega l}.
\end{equation}

\section{Results}\label{sec:mr}

In this section, we present a selection of our results concerning the absorption and scattering cross sections of massless test scalar waves in the background of Hayward spacetimes. We also compare our numerical results for the Hayward geometry with those obtained in the RN case.

\subsection{Numerical method}

We numerically solve Eq.~\eqref{RE_TC} from very close to the event horizon, i.e., $r_{\rm{initial}} = 1.001r_{+}$, to a region very far from the BH, typically chosen as $r_{\rm{\infty}} = 10^{3}M$. We then match the numerical solutions with the appropriated boundary conditions given by Eq.~\eqref{BC} and compute the reflection coefficient. Moreover, to calculate the absorption and scattering cross sections, we need to perform sums on the angular momentum of the scalar wave. For the absorption case, we typically set $l = 6$, while for the scattering case, we consider $l = 20$. Furthermore, the differential SCS has poor convergence for small values of the scattering angle. We improve the series convergence in this limit using the numerical method developed in Refs.~\cite{YRW1954,DDL2006}.

\begin{figure}[!htbp]
\begin{centering}
    \includegraphics[width=1.0\columnwidth]{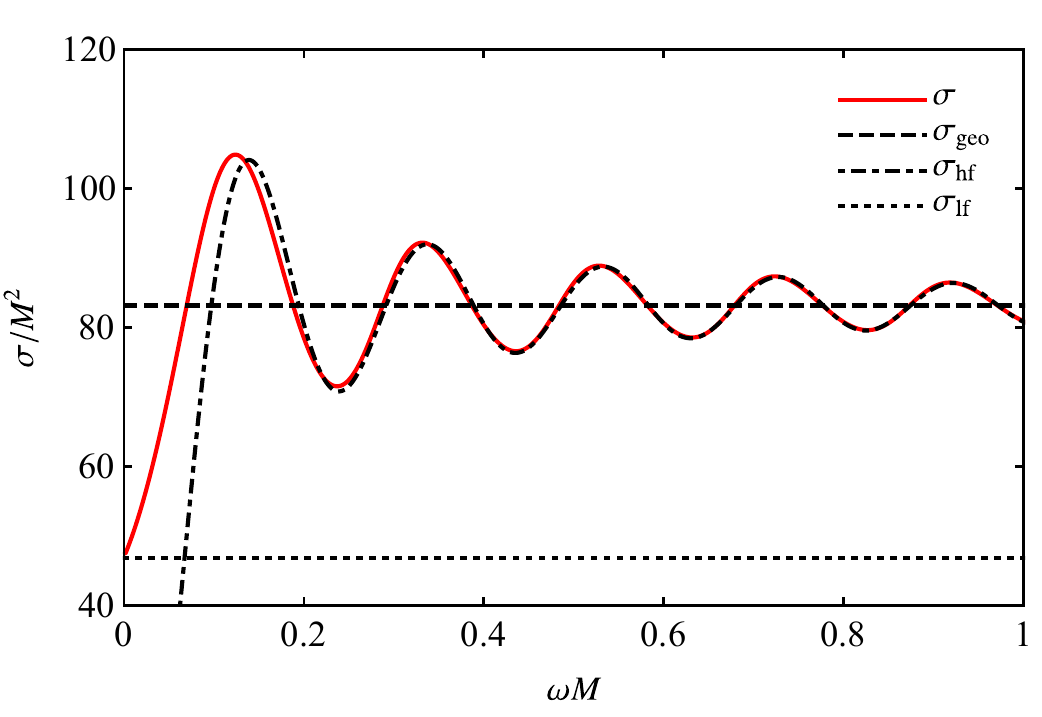}
    \caption{Comparison between the analytical approximations and numerical results for the total ACS of massless scalar fields of a Hayward RBH with $\alpha = 0.8$, as a function of $\omega M$.}
    \label{tacscomp}
\end{centering}
\end{figure}

In Fig.~\ref{tacscomp}, we compare our numerical results for the total ACS of massless scalar waves in the Hayward spacetime with some approximations. We can see that in the low-frequency regime, the total ACS tends to the event horizon area, namely
\begin{equation}
\sigma_{\rm{lf}} = 4\pi r_{+}^{2},
\end{equation}
as expected~\cite{DGM1997,H2001}. Moreover, in the high-frequency regime, our numerical results oscillate around the GCS [cf. Eq.~\eqref{GCS}] and the oscillatory pattern is well described by the sinc approximation [cf. Eq.~\eqref{SINC}], even for moderate frequency values.

\begin{figure}[!htbp]
\begin{centering}
    \includegraphics[width=1.0\columnwidth]{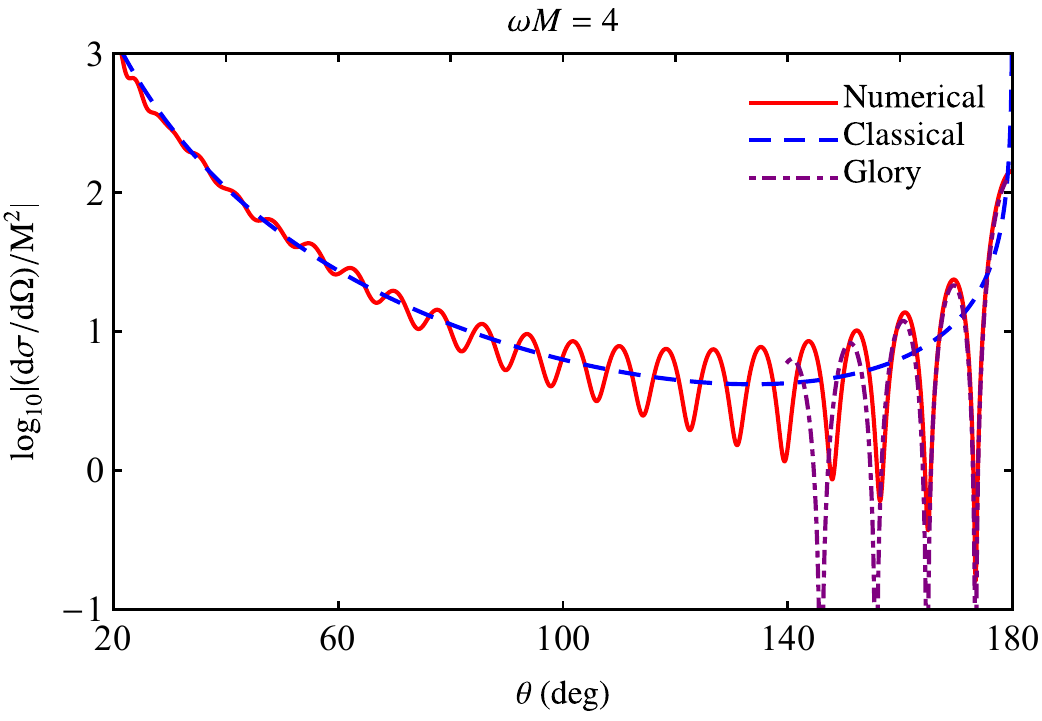}
    \caption{Comparison between the numerical and the approximate analytical results for the differential SCS of massless scalar fields in a Hayward RBH with $\alpha = 0.8$, as a function of $\theta$. Here we consider $\omega M = 4$.}
    \label{dscscomp}
\end{centering}
\end{figure}

Analogously, in Fig.~\ref{dscscomp}, we compare our numerical results for the differential SCS of massless scalar waves in the Hayward spacetime with some approximations. We can observe that the differential SCS oscillates around the classical differential SCS [cf. Eq.~\eqref{CSCS}] and the oscillatory pattern is well described by the glory approximation [cf. Eq.~\eqref{glory}] near the backward direction. We have, therefore, obtained excellent agreement between our numerical results and the approximated analytical ones.

\subsection{Massless scalar absorption}

\begin{figure}[!htbp]
\begin{centering}
    \includegraphics[width=1.0\columnwidth]{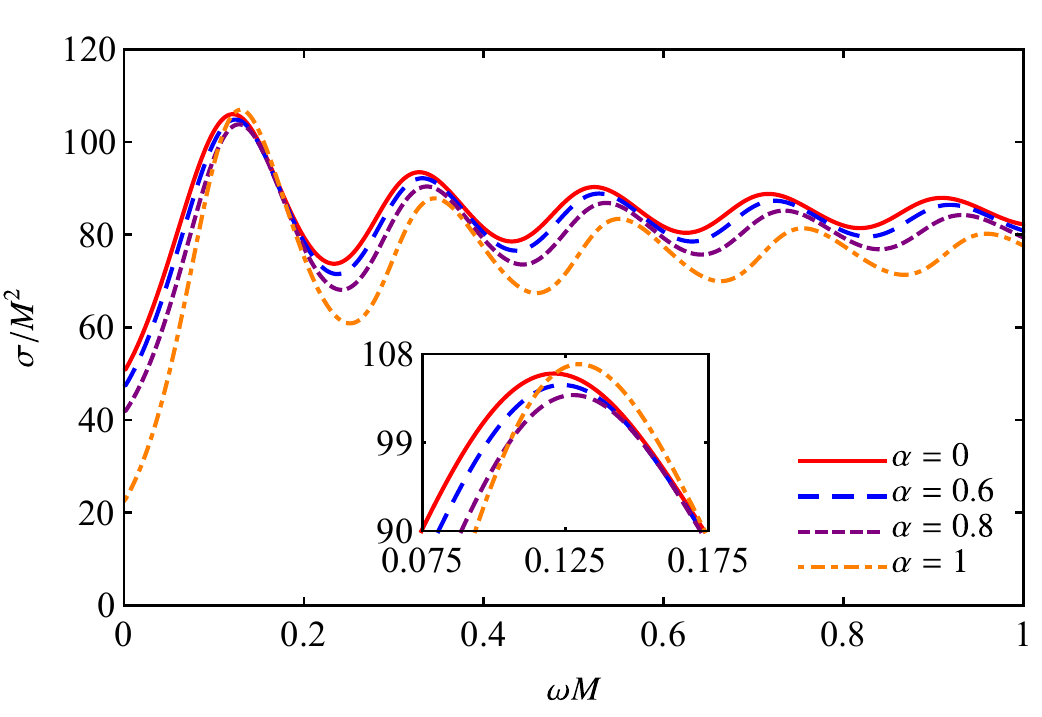}
    \includegraphics[width=1.0\columnwidth]{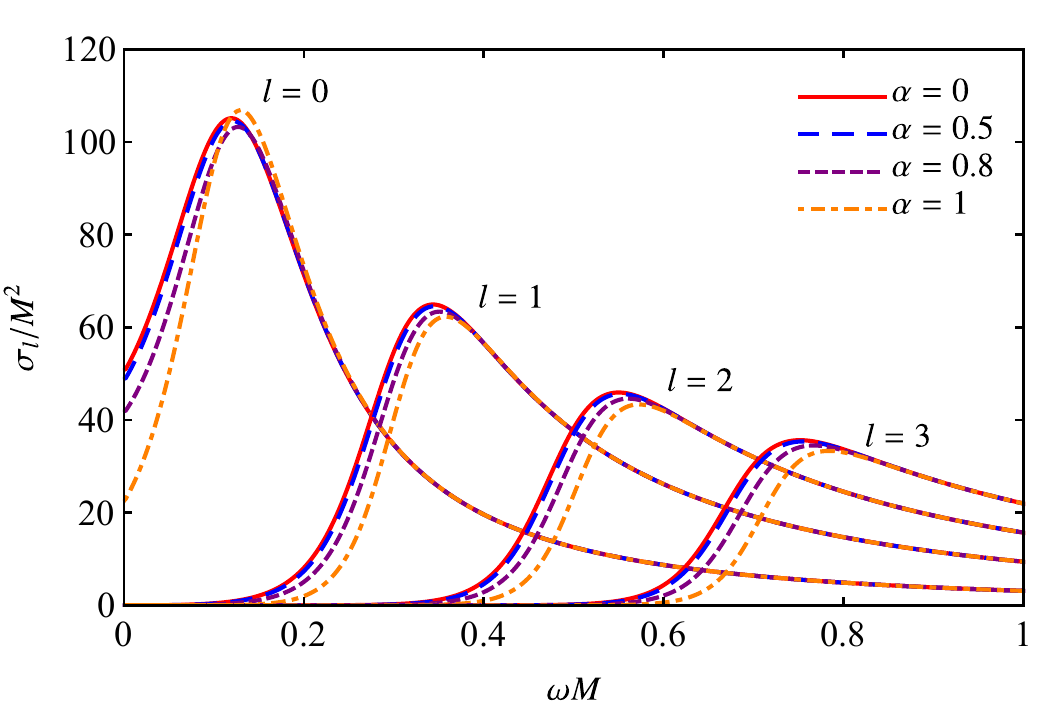}
    \caption{Total (top panel) and partial (bottom panel) ACSs of massless scalar fields in the Hayward spacetime, as functions of $\omega M$, considering different values of $l$ and $\alpha$. The inset in the top panel helps us to visualize the behavior of the total ACS near the first peak.}
    \label{tacsdifq}
\end{centering}
\end{figure}

In Fig.~\ref{tacsdifq}, we show the partial and total ACSs of scalar waves in Hayward spacetimes. As we can see, the total ACS typically decreases as we increase the BH charge. However, for $\alpha^{\rm{H}} \gtrsim 0.9658$, the first peak of the total ACS can be larger than in the Schwarzschild case ($\alpha = 0$). This feature can be related to the effective potential. As discussed in Sec.~\ref{subsec:msf}, the height of the potential barrier decreases as we enhance the values of $\alpha$ for $l = 0$. Therefore, massless scalar waves with $l = 0$ are more absorbed in the background of highly charged Hayward spacetimes, in contrast to what happens for $l \geq 1$. 

\begin{figure}[!htbp]
\begin{centering}
    \includegraphics[width=1.0\columnwidth]{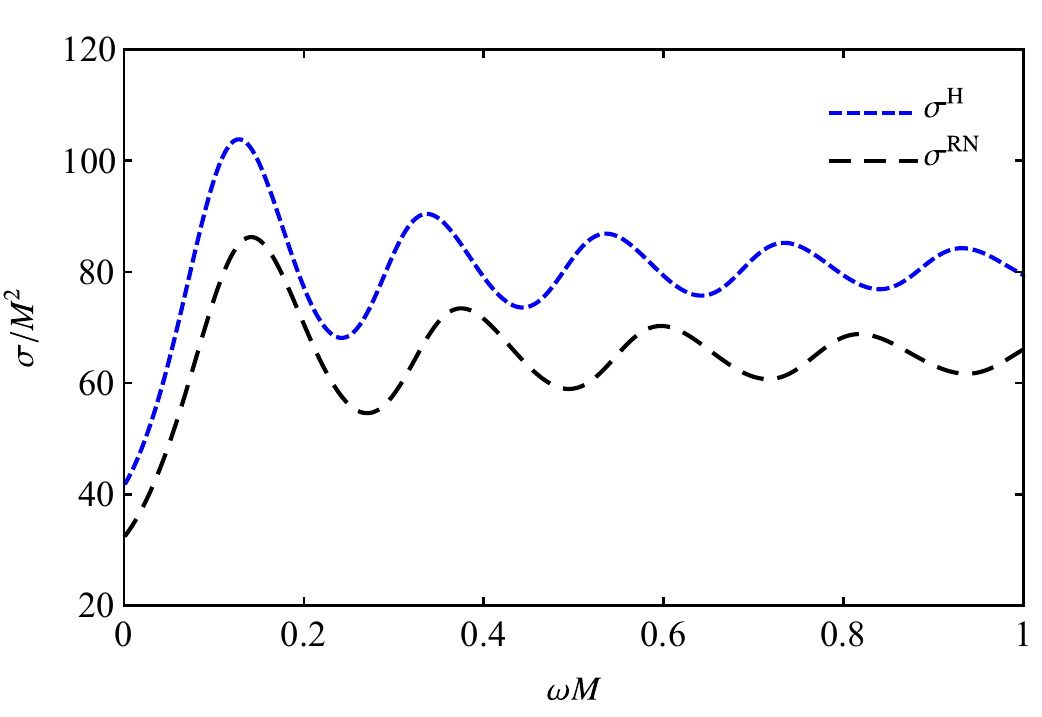}
    \caption{Comparison between the total ACSs, plotted as functions of $\omega M$, of massless scalar fields in Hayward and RN spacetimes with $\alpha =  0.8$.}
    \label{tacshdrn}
\end{centering}
\end{figure}

In Fig.~\ref{tacshdrn}, we compare the total ACSs of massless scalar waves in Hayward and RN spacetimes. As we can see, for the same value of the normalized charge, the total ACS in the Hayward RBH is typically larger than the corresponding one in the RN case. This result is consistent with the analysis presented in Sec.~\ref{subsec:msf}, since the effective potential of the RN BH is always greater than that of the Hayward RBH for the same values of $\alpha$.

\subsection{Massless scalar scattering}

\begin{figure}[!htbp]
\begin{centering}
    \includegraphics[width=1.0\columnwidth]{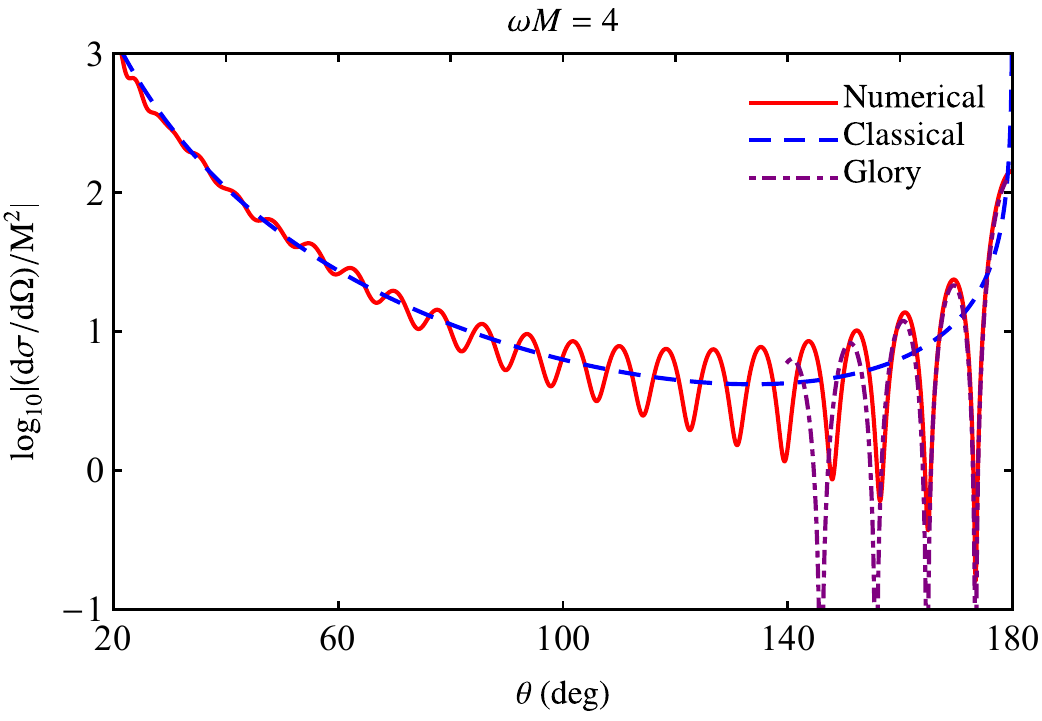}
    \caption{Differential SCSs of massless scalar fields in the Hayward spacetime, as a function of $\theta$, considering distinct values of $\alpha$, for $\omega M = 2$. The inset helps to visualize the differences in the interference fringe widths.}
    \label{dscs}
\end{centering}
\end{figure}

In Fig.~\ref{dscs}, we show a selection of our results for the scalar differential SCS in Hayward spacetimes. We can observe that the interference fringe widths get wider as we increase the BH charge, in agreement with the glory approximation~\cite{PLC2022,JBC2022,MLC2020-3}. We also notice that, for small scattering angles, the contributions of the BH charge are negligible, as stablished by Eq.~\eqref{CSCSweakHD}.

\begin{figure}[!htbp]
\begin{centering}
    \includegraphics[width=1.0\columnwidth]{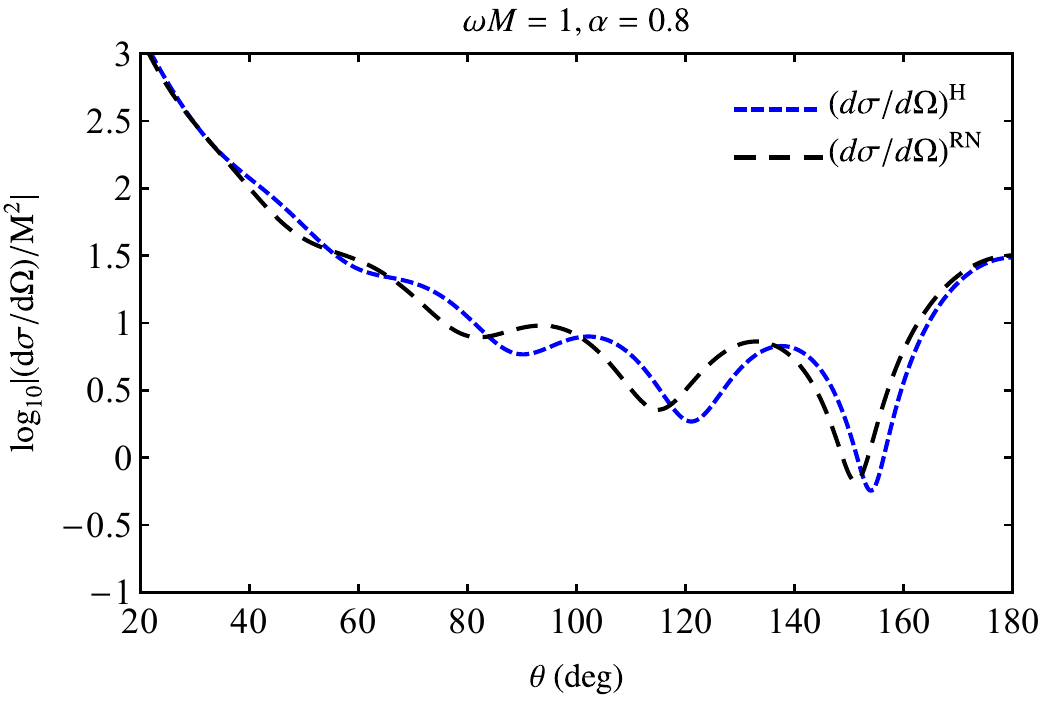}
    \caption{Comparison between the differential SCSs of massless scalar fields in Hayward and RN spacetimes with $\alpha =  0.8$, as a function of $\theta$, for $\omega M = 1$.}
    \label{dscshdrn}
\end{centering}
\end{figure}

A comparison between the scattering spectra of Hayward and RN BHs is presented in Fig.~\ref{dscshdrn}. For the same $\alpha$ values, the interference fringe widths for the RN spacetime are typically larger than  those in the Hayward corresponding case.

\subsection{Mimicking standard BHs}

We have also searched for situations in which the absorption and scattering spectra of Hayward and RN BHs can be very similar. Regarding configurations for which the ACSs are similar, we seek for the values of $\alpha$ that satisfy the following condition: $b_{c}^{\rm{H}} = b_{c}^{\rm{RN}}$. On the other hand, in the scattering case, we consider the values of the normalized charges for which the impact parameter of backscattered light rays matches, i.e., $b_{g}^{\rm{H}} = b_{g}^{\rm{RN}}$.

\begin{figure}[!htbp]
\begin{centering}
    \includegraphics[width=1.0\columnwidth]{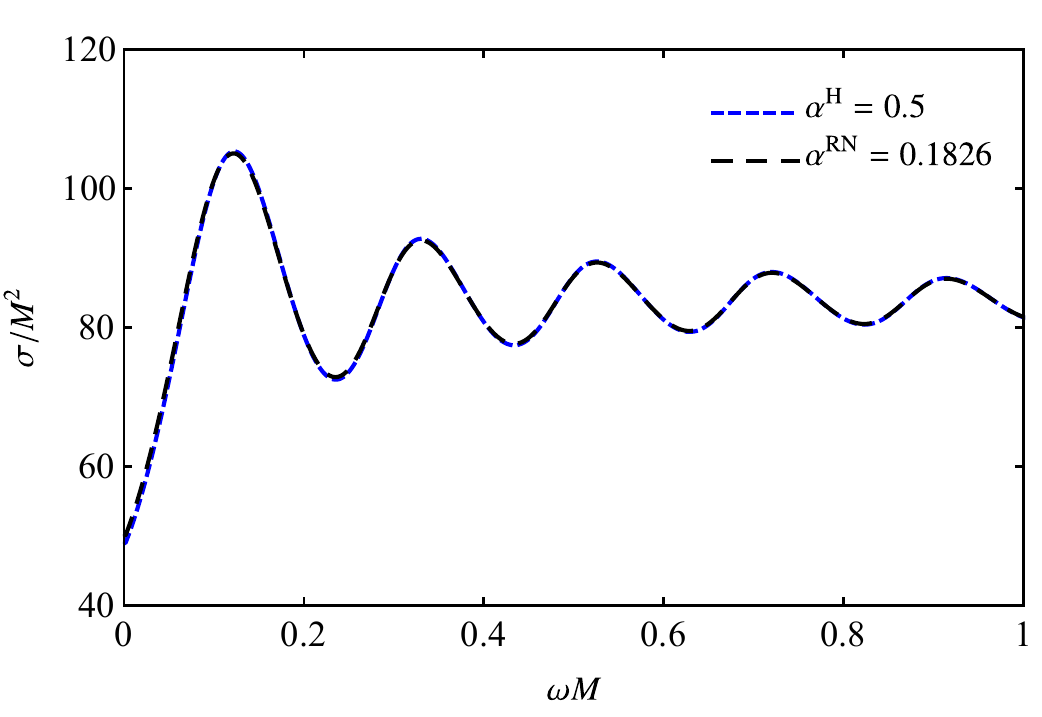}
    \includegraphics[width=1.0\columnwidth]{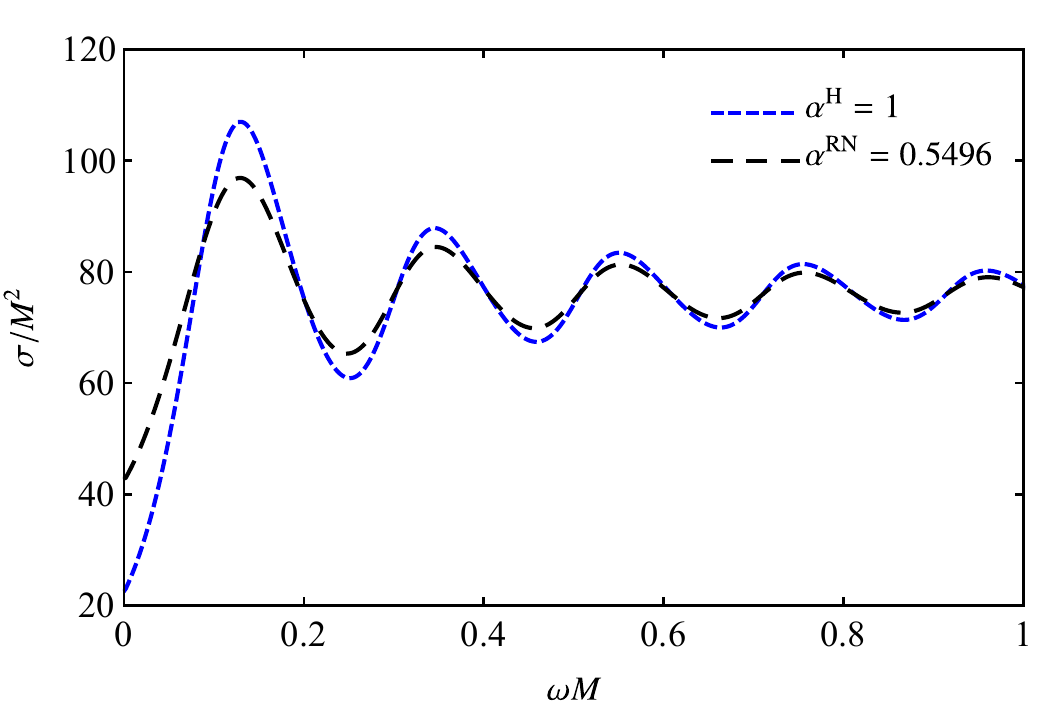}
    \caption{Comparison between the total ACSs of massless scalar fields in Hayward and RN spacetimes, as a function of $\omega M$. We have chosen $\alpha^{\rm{H}} = 0.5$ and $\alpha^{\rm{RN}} = 0.1826$, in the top panel; as well as $\alpha^{\rm{H}} = 1$ and $\alpha^{\rm{RN}} = 0.5496$, in the bottom panel.}
    \label{sacs}
\end{centering}
\end{figure}

A situation for which the ACSs of Hayward and RN BHs basically coincide is exhibited in the top panel of Fig.~\ref{sacs}. Indeed, we can find values of the charge for which the total ACSs can be very similar in the whole frequency range, as long as we consider low-to-moderate values of the normalized charges. 

\begin{figure}[!htbp]
\begin{centering}
    \includegraphics[width=1.0\columnwidth]{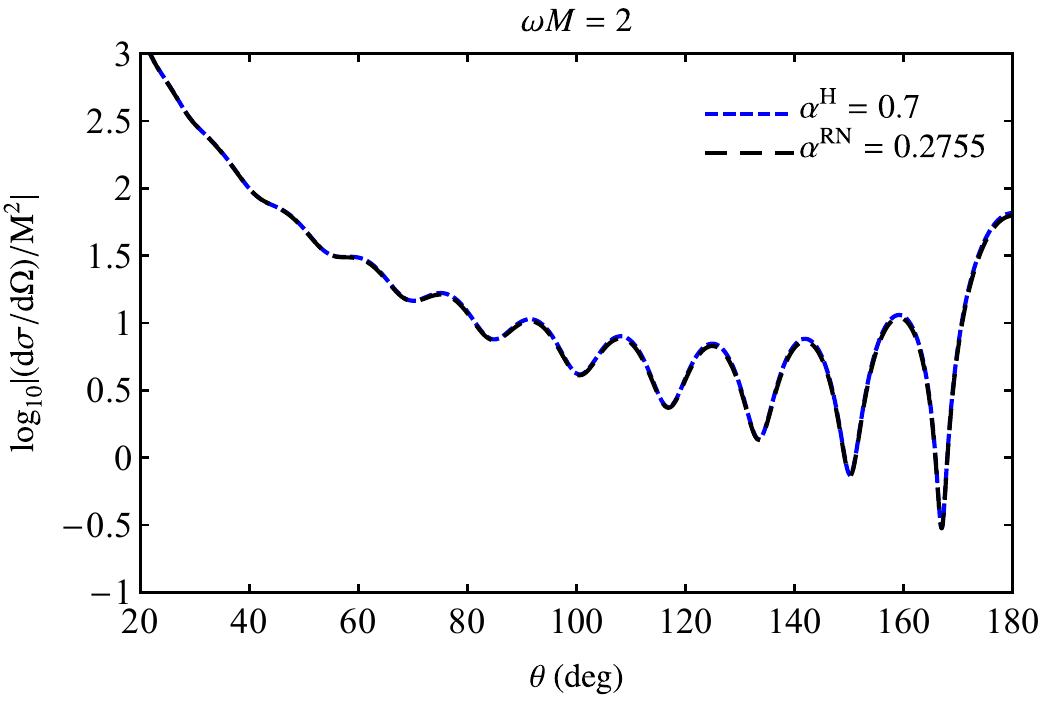}
    \includegraphics[width=1.0\columnwidth]{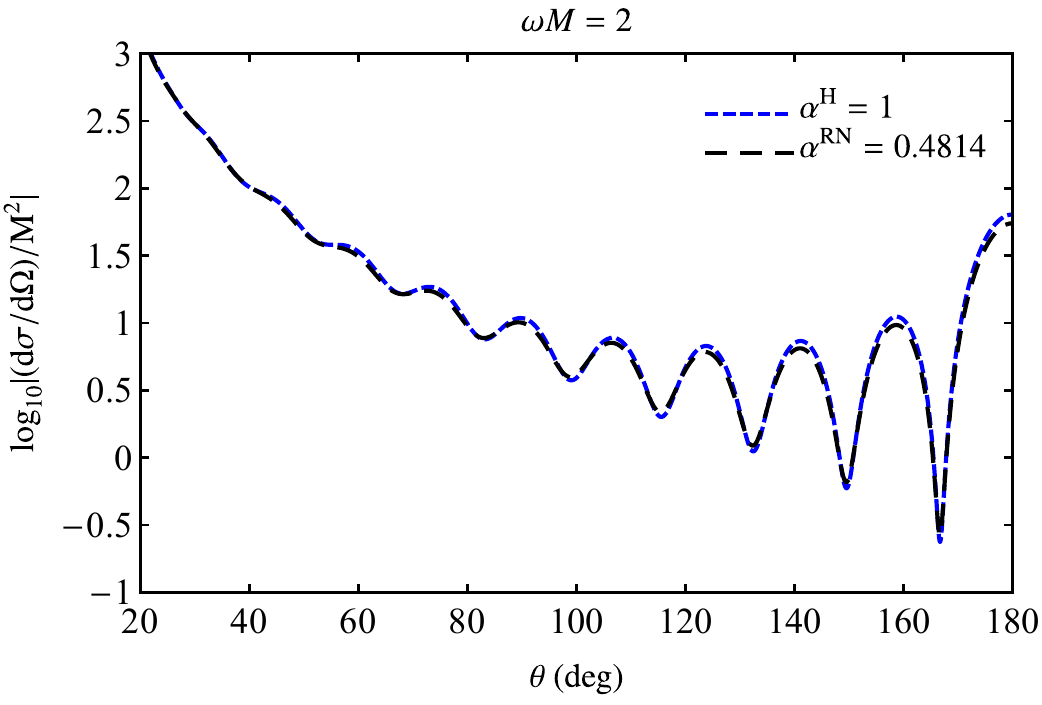}
    \caption{Comparison between the differential SCSs of massless scalar fields in Hayward and RN spacetimes, as a function of $\theta$, for $\omega M = 2$. We have chosen $(\alpha^{\rm{H}}, \alpha^{\rm{RN}}) = (0.7, 0.2755)$ (top panel), as well as $(\alpha^{\rm{H}}, \alpha^{\rm{RN}}) = (1, 0.4814)$ (bottom panel).}
    \label{sdscsa}
\end{centering}
\end{figure}

We exhibit, in the top panel of Fig.~\ref{sdscsa}, a situation for which the SCSs of Hayward and RN BHs basically coincide.
In Fig.~\ref{sdscsa}, we show the comparison between the differential SCSs in Hayward and RN spacetimes. 
For low-to-moderate values of the normalized charges, we can find configurations for which the SCSs are very similar. 
As we increase the BH charge,  keeping $b_{g}^{\rm{H}} = b_{g}^{\rm{RN}}$, the oscillatory profile remains similar, but the differences become more evident. 

\section{Final remarks}\label{sec:fr}

We have investigated the absorption and scattering spectra of massless scalar fields in the background of the Hayward RBH solution. We have compared our numerical results obtained for arbitrary values of the frequency and scattering angle of the scalar wave with some analytical approximations, showing that they are in excellent agreement. We have noticed that the interference fringe widths get wider as we increase the RBH charge or decrease the frequency.

Concerning the absorption properties, we have obtained that, although the total ACS typically decreases as we enhance the RBH charge, the first peak (local maximum related to the monopole mode, $l=0$) of the ACS in the Hayward spacetime can be larger than in the Schwarzschild case, for $\alpha^{\rm{H}} \gtrsim 0.9658$, while the local maxima related to $l>0$ are smaller than those of Schwarzschild, assuming any (positive) value of magnetic charge. This distinctive feature may be related to the behavior of the effective potential.

We also have noticed that it is possible to find configurations for which the absorption and scattering of massless scalar waves in Hayward and RN geometries are very similar. These similarities can be found for general values of the frequency and scattering angle of the scalar wave but are constrained to low-to-moderate values of the BH charges. Our results reinforce that regular and singular BHs can have very similar absorption and scattering properties under certain circumstances, but we might be able to distinguish between them in other scenarios~\cite{JRV2023}.

It is worth mentioning that recently the absorption and scattering properties of massless test scalar fields in Hayward RBH spacetimes were addressed in Ref.~\cite{WW2022}, but the results obtained there were not sound~\cite{PLC2023b}. Here we provided the correct results.

\begin{acknowledgments}

We are grateful to Funda\c{c}\~ao Amaz\^onia de Amparo a Estudos e Pesquisas (FAPESPA), Conselho Nacional de Desenvolvimento Cient\'ifico e Tecnol\'ogico (CNPq) and Coordena\c{c}\~ao de Aperfei\c{c}oamento de Pessoal de N\'ivel Superior (CAPES) -- Finance Code 001, from Brazil, for partial financial support. MP and LC thank the University of Sheffield, in England, and University of Aveiro, in Portugal, respectively, for the kind hospitality. LL would like to acknowledge IFPA -- Campus Altamira for the support. This work has further been supported by the European Union's Horizon 2020 research and innovation (RISE) programme H2020-MSCA-RISE-2017 Grant No. FunFiCO-777740 and by the European Horizon Europe staff exchange (SE) programme HORIZON-MSCA-2021-SE-01 Grant No. NewFunFiCO-101086251.

\end{acknowledgments}

\end{document}